\documentclass{elsart}
\usepackage{epsfig}
\begin{document}
\begin{frontmatter}
\title{Electro-Optical Detection of Charged Particles \thanksref{doe}}
\thanks[doe]{Work supported in part by the U.S. Department of Energy under
Contract No. DE-AC02-98CH10886}
\author[bnl]{Y.K. Semertzidis},
\author[bnl]{V. Castillo}, 
\author[msu]{L. Kowalski},
\author[pitt]{D.E. Kraus}, 
\author[bnl]{R. Larsen},
\author[bnl]{D.M. Lazarus\thanksref{tel}},
\author[bnl]{B. Magurno\thanksref{gone}}, 
\author[bnl]{D. Nikas},
\author[bnl]{C. Ozben},
\author[bnl]{T. Srinivasan-Rao}, 
\author[bnl]{T. Tsang}
\address[bnl]{Brookhaven National Laboratory, Upton, NY 11973, U.S.A.}
\address[msu]{Montclair State University, Upper Montclair, NJ 07043 U.S.A.}
\address[pitt]{University of Pittsburgh, Pittsburgh PA, 15260 U.S.A.}
\thanks[tel]{Corresponding author. Tel.: +1-631-344-3959;fax: +1-631-344-5954\\
{\it E-mail address}: lazarus@sun2.bnl.gov}
\thanks[gone]{Deceased.}
\begin{abstract}
We have made the first observation of a charged particle beam by means of its 
electro-optical effect on the polarization of laser light in a LiNbO$_3$ 
crystal. The modulation of the laser light during the passage of a pulsed 
electron beam was observed using a fast photodiode and a digital oscilloscope. 
The fastest rise time measured, 120 ps, was obtained in the single shot mode 
and was limited by the bandwidth of the oscilloscope and the associated 
electronics. This technology holds promise for detectors of greatly improved 
spatial and temporal resolution for single relativistic charged particles as 
well as particle beams.
\end{abstract}
\end{frontmatter}

\section{Introduction}
An effort has been initiated to develop an ultra-fast charged particle detector
based on the birefringence, i.e. the anisotropy of the index of refraction, 
produced in an optical medium by the electric field of relativistic charged 
particles~\cite{gls}. The anisotropy of the index of refraction implies 
different velocities of light propagation for orthogonal components of 
polarization, resulting in a phase difference between them which will
produce an elliptical polarization in light that had initially been linearly 
polarized. 
The ellipticity is defined as the ratio of the semi-minor to the semi-major 
axis of the ellipse traced by the electric field vector projected on a plane 
perpendicular to the direction of propagation. The electro-optical effect in 
amorphous optical 
media is known as the Kerr effect~\cite{kerr} and is quadratic in the electric 
field $E$, $\phi = 2 \pi K E^2 d$, where $\phi$ is the phase difference induced
between orthogonal components of the polarized light. $K$ is the Kerr 
coefficient and $d$ is the length of the electric field region experienced by 
the material.  

In uniaxial crystals, the induced ellipticity is linear in the  E-field, and 
is known as the Pockels effect~\cite{yariv}. The difference in indices of 
refraction along orthogonal axes, $\Delta n=PE$ where P is the Pockels
constant and E is the electric field. For LiNbO$_3$, the Pockels constant 
$P=31$~pm/V. The induced phase delay can be written as
$\phi=2\pi\Delta nd/\lambda = \pi(V/V_\pi)$ with $V$ the applied voltage
and $V_\pi$ the voltage required for producing a phase shift between orthogonal
components of polarization equal to $\pi$ radians. The maximum ellipticity of 
$\pi/\sqrt{2}$ is induced when the polarization is at $45^\circ$ with respect 
to the applied electric field direction.

Devices based on the Pockels effect have long been employed in the detection 
of fast transient electrical signals as well as for modulation of optical
signals~\cite{early}.

\section{Apparatus}
The 45 MeV kinetic energy electron beam at the Brookhaven National Laboratory 
Accelerator Test Facility, containing up to 1 nC in a diameter of approximately
1 mm with 10 ps duration and a repetition rate of 1.5~Hz, was scanned over a
range of distances of several millimeters from the crystal.

The passage of beam bunches was detected through the effect of their electric 
field on the polarization of the light in a commercially available LiNbO$_3$ 
electro-optical modulator crystal.~\cite{utp}. The optics setup is depicted 
schematically in Fig.1 The LiNbO$_3$ crystal was coupled to two 40 m long 
polarization maintaining fibers of 4 microns core diameter. The first of the 
fibers propagated polarized light of wavelength $\lambda = 1.32 \mu${\rm m} 
from a 20 mW CW Amoco laser operating at approximately 10 mW and the second 
fiber returned the transmitted light to a quarter wave plate which transformed 
the induced ellipticity to a rotation of the initially linear polarization 
which had previously been nearly extinguished by the analyzer (crossed
polarizer) in the absence of the electron beam. The resulting modulation of the
transmitted laser light was detected by a photodiode and pre-amplifier of 45 
GHz bandwidth~\cite{newfocus} which was followed by a 10 GHz amplifier with 32
dB gain. The output signals were digitized in a 7 GHz oscilloscope~\cite{hp} 
and stored in memory.
\begin{figure}
%\begin{center}
\vspace{5mm}
\epsfig{file=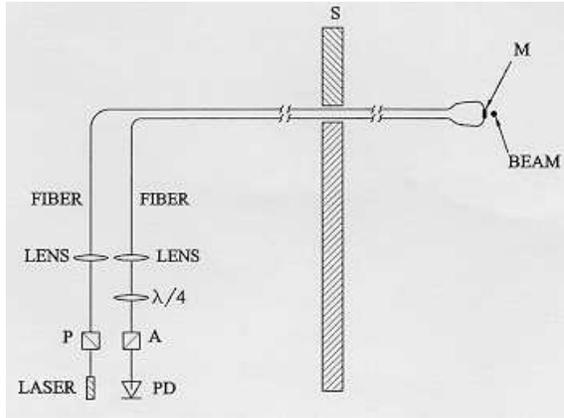, width=75mm}
%\vspace*{-3cm}
%\end{center}
\caption{The experimental setup for detecting a charged particle beam. The 
LiNbO$_3$ crystal (M) was located in vacuum several mm from the beam position 
which could be varied over several cm. The beam direction was perpendicular 
to the plane of the page. The positions of the polarization maintaining fibers,
polarizer (P), lenses, $1/4$ wave plate ($\lambda/4$), analyzer (A), shield 
wall (S) and photodiode detector(PD) are schematically indicated.}
\label{setup}
\end{figure}

The crystal modulator package had been modified to allow for the passage of the
electron beam without hitting the housing. The transmitted light intensity, I, 
exiting the analyzer, A, is given by
\begin{equation}
I = I_0 [ \sigma^2 + (\alpha + \epsilon(t))^2 ] \approx I_0 [ \sigma^2 +
\alpha^2 + 2 \alpha \epsilon(t) ] \label{eq:intensity} 
\end{equation}
\noindent{with} $\epsilon$ the induced ellipticity, and $\alpha$ the 
misalignment angle. $\sigma^2$ is the minimum possible ratio of ($I/I_0$) 
when ~$\alpha$ and ~$\epsilon$ are equal to zero and is defined as the 
extinction. 
A finite $\alpha$ can be used to linearize and amplify the effect on 
$\epsilon$. It is apparent from Eq.~\ref{eq:intensity}, the time dependent part
of the light signal can be made positive or negative depending on the sign of 
$\alpha$ relative to $\epsilon$.

A bunched charged particle beam creates an electric field, $E$, at a distance 
$r$ that is well approximated by Coulomb's Law multiplied by $\gamma$, the 
relativistic Lorentz factor, and $N_e$ the number of electron charges in the 
beam bunch for a minimum distance $r_0$ much greater than the dimensions of the
beam bunch.
\begin{equation}
E=\gamma N_e{q\over 4\pi\epsilon_0 r_0^2 }
\end{equation}
The electric field is present for $\Delta t=r_0/{\gamma c}$ with $r_0$ the
effective range of the electric field, $E$, and $c$ the beam velocity. Laser 
light travels a distance 
\begin{equation}
\Delta l = (c/n)\Delta t = r_0/{n\gamma}
\end{equation}
in the crystal during the time that the electric field is present. This
determines the resolving time for this technique. 
Since the beam was present in $10$ ps bunches, the photon detection system 
had insufficient high frequency bandwidth to avoid signal attenuation but the
high intensity, $\approx 10^{10}$ electrons/ bunch, provided sufficient signal
amplitude for detection.

The electric field sensitivity of the LiNO$_3$ crystal was determined from the 
modulator specification provided by the manufacturer for the dependence of the 
phase difference on an applied high frequency voltage. For a phase difference 
$\pi$, the LiNbO$_3$ crystal had $V_{\pi}=5.7$~V with an electrode separation 
of $15\mu$m and a length of $l=1.5$~cm. The integral 
\begin{equation}
(\int E\,dl)_{\pi}= \int 5.7V/(15\times 10^{-6}\,{\rm m})\,dl=5700~V 
\end{equation}
produces the maximum ellipticity of $\pi/\sqrt{2}$.

The integral of the electric field of the particle bunch over the crystal 
dimensions for a beam bunch located at the mid-plane orthogonal to the length 
of the crystal at a distance $r_0$ is
\begin{equation}
\int Edl\approx\frac{N_e q}{4\pi\kappa\epsilon_0 n r_0}
\end{equation}
where $\kappa$ is the dielectric constant and n is the index of refraction of 
the LiNbO$_3$. For $r_0=5$~mm, $n=2.2$ and $\kappa=32$ for the electric 
field aligned with the r$_{33}$ axis, $\int Edl=4.1\times 10^{-9} N_e$~V 
which with Equation 4 yields an ellipticity
\begin{equation}
\epsilon=\frac{\int Edl}{(\int Edl)_{\pi}}\times{\pi}/\sqrt{2}=
7.2\times 10^{-13} N_e
\end{equation}
The signal to noise ratio for a measurement of this type that is photon 
statistics limited is given by 
\begin{equation}
SNR =\epsilon \sqrt{P T q_p \over 2 h\nu} 
\end{equation}
where $P$ is the laser power, $T$ the time resolution or inverse of the 
detection system bandwidth, $q_p$ the quantum efficiency of the photodiode,
and $h\nu$ the laser photon energy. Excellent time resolution results in
enhanced sensitivity, which can be essential for the detection of extremely 
short pulses, but it will also increase white noise acceptance due to the 
increased bandwidth.\\

\section{Experimental Results}
Measurements were made in the single shot mode using the 7 GHz oscilloscope.
The extinction was approximately $2\times 10^{-3}$.
With $P=10$~mW, $T=100$~ps, $q_p=0.8$ and $h\nu=0.9$~eV, Eqs.6 and 7
determine the required number of electrons in the beam for $SNR=1$, 
$N_e\approx 8\times 10^8$ for detection of a single beam bunch. However, the
7 GHz bandwidth limits the sensitivity to the inherently much faster signal
by as much as factor of five implying $\approx 4\times 10^9$ singly charged 
particles are required for detection of a single ATF beam bunch. The ATF beam 
of $N_e=6.3\times 10^9$ per bunch was sufficient to generate a detectable 
modulation. 

The polarization dependent signal is displayed in Fig. 2 (solid 
curve). Changing the sign of $\alpha$ (see Eq.~\ref{eq:intensity}) caused the 
sign of the signal to change whereas the amplitude was unaffected. The maximum 
modulation of the light intensity was $\approx 9\%$ of its DC level.
\begin{figure}
%\begin{center}
\vspace{10mm}
\epsfig{file=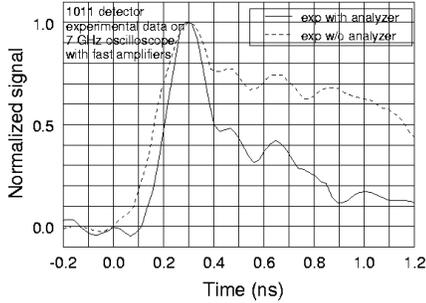, width=60mm,angle=-90}
%\vspace*{-1.1cm}
%\end{center}
\caption{The polarization dependent signal (solid line).  The electron beam 
was $\approx 0.5$~cm from the crystal.  The polarization independent
signal is indicated by the dashed line.}
\label{signal1}
\end{figure}

A  negative signal due to the attenuation of light was observed when the 
LiNO$_3$ crystal intercepted the beam. It was found to be polarization 
independent and has a significantly longer decay time constant than the 
polarization dependent signal. A similar effect was found in poled quartz
fibers and polarization maintaining fibers.

\section{Discussion}
Although it presently appears possible to construct a beam profile detector 
with excellent time resolution based on parallel rows of LiNbO$_3$ crystals
, a more formidable challenge would be to achieve single particle detection 
with this technology. With a high power pulsed laser of $10^9$~W, i.e. 10 mJ 
for 10 ps and a transient digitizer of greater bandwidth, e.g. 60 GHz, a 
detector constructed of parallel rows of electro-optical crystals with a 
separation of $100\mu$m and therefore an effective average distance, 
$r_0\approx 25\mu$m between a single charged particle trajectory and the 
nearest fiber can be assumed. Equations 5 and 6 would then give an ellipticity 
$\epsilon=1.4\times 10^{-10}$ and Equation 7 gives 
$SNR\approx 2.3\times 10^{-2}$ for single particle detection with LiNiO$_3$.

In addition, use of a high power pulsed laser would preclude the detection of 
randomly occurring events because of the inability to trigger the laser. 
However, this would be ameliorated somewhat at colliding beams machines where 
the beams are tightly bunched and interact periodically to maximize the 
luminosity. 
  
The small size and high cost of the LiNbO$_3$ crystals has prompted a search
for materials of larger electro-optical coefficients and lower dielectric
constants. We have measured the Kerr coefficients of optical fibers and quartz 
rods to be $10^{-4}-10^{-3}$~pm/V$^2$. Organic liquids such as nitrobenzene 
have relatively large Kerr coefficients and could be used in hollow glass 
fibers of micron dimensions. Although these could lead to inexpensive devices, 
the Kerr coefficients are less than 2.5 pm/V$^2$ which is far too small for 
sensitivity to singly charged particles. 

Electrically poled optical fibers have been shown to produce relatively large 
electro-optical effects~\cite{pole}. At sensitivities as high as 0.1 nm/V$^2$
they may be adequate for charged particle beam diagnostics but it is unlikely 
that it will be sufficiently sensitive for single particle detection. 

DAST~(4-dimethylamino-N-methyl-4-stilbazolium tosylate) crystals~\cite{dast}
have been found to have large electro-optic coefficients, from $47\pm 8$~pm/V 
at $\lambda=1535$~nm to $92\pm 9$~pm/V at $\lambda=720$~nm as well as a small 
dielectric constant, $\kappa=5.2$ compared to 30.8 pm/V and $\kappa=32$ for 
LiNbO$_3$. This implies a factor of 20 greater sensitivity than
LiNbO$_3$. If DAST or yet to be developed electro-optical materials
can yield adequate sensitivity for single particle detection, a new generation
of ultrafast fine grained charged particle detectors may be at hand.  

\section{Acknowledgements}

We wish to acknowledge the support and encouragement of Xiejie Wang, Ilan
Ben-Zvi, Vitaly Yakimenko, John Miller, Howard Gordon, Mike Murtagh and Tom 
Kirk. The efforts of Victor Usack and the contributions of our late colleague 
Ben Magurno were essential to our progress.


\begin{thebibliography}{9}
\bibitem{hp}  The digitizing oscilloscope used was a Tektronix SDC 5000 of 
7~GHz bandwidth.
\end{thebibliography}
\end{document}